\documentclass[12pt]{article}
\usepackage{amsmath}
\usepackage{graphicx}

\newsavebox{\SKpath}
\sbox{\SKpath}{\includegraphics[width=8.8cm]{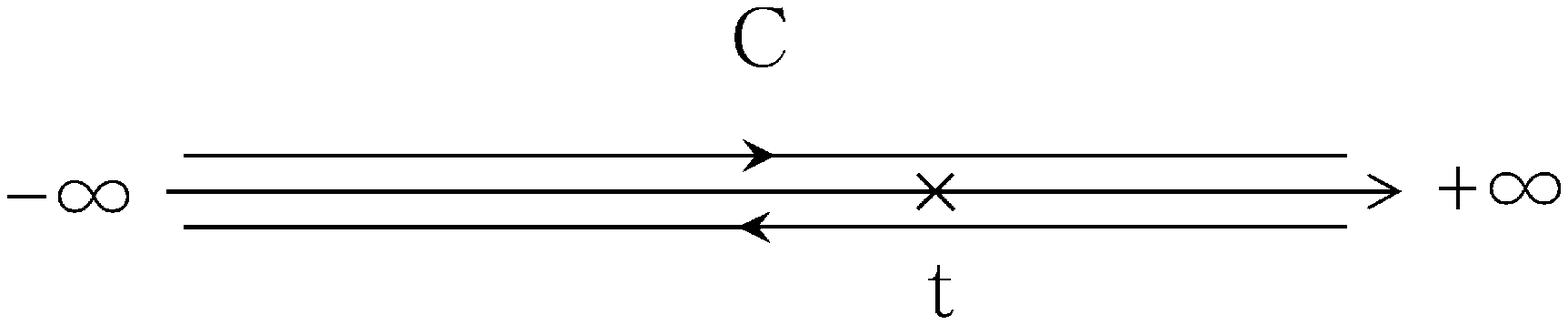}} 
\newlength{\SKpathl}
\def\papertitlepage{\baselineskip 3.5ex\thispagestyle{empty}}
\def\preprinumber#1#2{\hfill\begin{minipage}{4.2cm} #1
        \par\noindent #2 \end{minipage}}
\allowdisplaybreaks[3]

\begin{document}

\papertitlepage
\setcounter{page}{0}
\preprinumber{KEK-TH-1539}{}
\baselineskip 0.8cm
\vspace*{2.0cm}

\begin{center}
{\Large\bf Soft Graviton effects on Gauge theories\\
 in de Sitter Space}
\end{center}

\begin{center}
Hiroyuki K{\sc itamoto}$^{1)}$\footnote{E-mail address: kitamoto@post.kek.jp} and
Yoshihisa K{\sc itazawa}$^{2),3)}$\footnote{E-mail address: kitazawa@post.kek.jp}\\
\vspace{5mm}
$^{1)}${\it Department of Physics and Astronomy\\
Seoul National University, 
Seoul 151-747, Korea}\\
$^{2)}${\it KEK Theory Center, 
Tsukuba, Ibaraki 305-0801, Japan}\\
$^{3)}${\it Department of Particle and Nuclear Physics\\
The Graduate University for Advanced Studies (Sokendai)\\ 
Tsukuba, Ibaraki 305-0801, Japan}
\end{center}

\vskip 5ex
\baselineskip = 2.5 ex

\begin{center}{\bf Abstract}\end{center}

We extend our investigation of soft graviton effects
on the microscopic dynamics of matter fields in de Sitter space.
We evaluate the quantum equation of motion in generic gauge theories.
We find that the Lorentz invariance can be respected and the
velocity of light is not renormalized at the one-loop level.
The gauge coupling constant is universally screened by soft gravitons and diminishes with time.
These features are in common with other four dimensional field theories with dimensionless couplings.
In particular the couplings scale with time with definite scaling exponents.
Although individual scaling exponents are gauge dependent, we argue that the relative scaling
exponents are gauge independent and should be observable.
We also mention soft graviton effects on cosmic microwave background.

\vspace*{\fill}
\noindent
June 2013

\newpage
%------------------------------------------------------------------------------------
\section{Introduction}
\setcounter{equation}{0}

The propagator of a massless and minimally coupled scalar field 
contains a de Sitter (dS) symmetry breaking term. 
This term is a direct consequence of the scale invariant spectrum at the super-horizon scale
and depends logarithmically on the scale of the Universe \cite{Vilenkin1982,Linde1982,Starobinsky1982}. 
Here we consider an exponentially expanding Universe which begins with a finite spatial extension.
In some field theoretic model in dS space, 
such an effect gives rise to growing time dependences to physical quantities through quantum corrections. 

By employing the Schwinger-Keldysh formalism \cite{Schwinger1961,Keldysh1964}, 
we can investigate interacting field theories in dS space perturbatively. 
The IR effects at each order manifest as the polynomials 
in the logarithm of the scale factor of the Universe: $\log a(\tau)$ \cite{Weinberg2005}. 
For example, such IR effects are seen in an interacting scalar field theory 
such as $\lambda\varphi^4$ theory \cite{Woodard2002,Onemli2004}. 
Let us consider the expectation value of the energy-momentum tensor. 
Here the leading IR effect at the $(n+1)$-loop level manifests as $(\lambda\log^2a(\tau))^n$. 
It indicates that 
the quantum effect may grow until it saturates when $\lambda\log^2a(\tau)\sim 1$. 
Actually, we can confirm the existence of such an equilibrium state 
in the stochastic approach \cite{Starobinsky1994,Woodard2005}. 

The necessary condition for the dS symmetry breaking is the existence of
a massless minimally coupled scalar field. 
We have investigated the IR effects of the non-linear sigma model where
the reparametrization symmetry of the target space ensures the existence of
massless minimally coupled scalar fields.
We have shown the cancellation of the leading IR logarithms to all orders \cite{Kitamoto2011}. 
For a general scalar field theory, 
we need to fine-tune the mass term to obtain a massless scalar field. 
On the other hand, 
the gravitational field contains massless and minimally coupled modes without the fine-tuning. 
In this regard, the gravitational field is an attractive candidate which induces the IR effects. 
In a similar way to $\lambda\varphi^4$ theory, the IR effects from gravity may grow 
up to certain values which are not suppressed by the gravitational constant. 
Such an effect may be relevant to understand the cosmological constant problem.

Although we cannot observe the super-horizon modes directly, 
it is possible that virtual gravitons of the super-horizon scale affect 
the local dynamics of the sub-horizon scale. 
In fact 
we have found that soft gravitons screen the couplings of $\phi^4$ and Yukawa interactions
 \cite{Kitamoto2012}. 
In this paper we investigate the identical effect with respect to gauge interactions. 
In addition to the $\phi^4$ and Yukawa interaction, 
the investigation of gauge theory is necessary
in order to understand the soft gravitation effects on the standard model of particle physics. 

The organization of this paper is as follows. 
In Section 2, we recall our quantization procedure of gravitational field on dS background. 
We identify the graviton modes which exhibit IR logarithm. 
In Section 3, we evaluate the quantum equation of motion with respect to gauge fields 
which are corrected by soft gravitons at the one-loop level. 
We find that the effective couplings of gauge interactions are screened by soft gravitons. 
In Section 4, we introduce a gauge parameter of the graviton propagator 
to investigate the gauge dependences of the results obtained in Section 3. 
By comparing the  gauge couplings with that of $\phi^4$ and Yukawa interactions,
we show that the relative scaling exponents of the dimensionless couplings are gauge independent and observable. 
We conclude in Section 5. We mention possible implications of our results on cosmic microwave background.

%------------------------------------------------------------------------------------
\section{Gravitational field in de Sitter space}
\setcounter{equation}{0}

In this section, we review the gravitational propagator in de Sitter (dS) space.  
In the Poincar\'{e} coordinate, the metric in dS space is
\begin{align}
ds^2=-dt^2+a^2(t)d{\bf x}^2,\hspace{1em}a(t)=e^{Ht}, 
\end{align}
where the dimension of dS space is taken as $D=4$ and $H$ is the Hubble constant. 
In the conformally flat coordinate,
\begin{align}
(g_{\mu\nu})_\text{dS}=a^2(\tau)\eta_{\mu\nu},\hspace{1em}a(\tau)=-\frac{1}{H\tau}. 
\end{align}
Here the conformal time $\tau\ (-\infty <\tau < 0)$ is related to the cosmic time $t$ as $\tau\equiv-\frac{1}{H}e^{-Ht}$. 
We assume that dS space begins at an initial time $t_i$ with a finite spacial extension.
After a sufficient exponential expansion, the dS space is well described locally by the above metric irrespective of the spacial topology. 
The metric is invariant under the scaling transformation: 
\begin{align}
\tau\to c\tau,\hspace{1em}x^i\to cx^i. 
\end{align}
It is a part of the $SO(1,4)$ dS symmetry.

In dealing with the quantum fluctuation on the dS background, 
we adopt the following parametrization: 
\begin{align}
g_{\mu\nu}=\Omega^2(x)\tilde{g}_{\mu\nu},\ \Omega(x)=a(\tau)e^{\kappa w(x)}, 
\label{para1}\end{align}
\begin{align}
\det \tilde{g}_{\mu\nu}=-1,\ \tilde{g}_{\mu\nu}=\eta_{\mu\rho}(e^{\kappa h(x)})^{\rho}_{~\nu}, 
\label{para2}\end{align}
where $\kappa$ is defined by the Newton's constant $G$ as $\kappa^2=16\pi G$. 
To satisfy (\ref{para2}), $h^{\mu}_{~\nu}$ is taken to be traceless
\begin{align}
h^{\mu}_{~\mu}=0. 
\label{para3}\end{align}
In this parametrization, the scalar density and the Ricci scalar are written as  
\begin{align}
\sqrt{-g}=\Omega^4,\hspace{1em}
R=\Omega^{-2}\tilde{R}-6\Omega^{-3}\tilde{g}^{\mu\nu}\nabla_\mu\partial_\nu\Omega, 
\label{com1,2}\end{align}
where $\tilde{R}$ is the Ricci scalar constructed from $\tilde{g}_{\mu\nu}$
\begin{align}
\tilde{R}=-\partial_\mu\partial_\nu\tilde{g}^{\mu\nu}
-\frac{1}{4}\tilde{g}^{\mu\nu}\tilde{g}^{\rho\sigma}\tilde{g}^{\alpha\beta}\partial_\mu\tilde{g}_{\rho\alpha}\partial_\nu\tilde{g}_{\sigma\beta}
+\frac{1}{2}\tilde{g}^{\mu\nu}\tilde{g}^{\rho\sigma}\tilde{g}^{\alpha\beta}\partial_\mu\tilde{g}_{\sigma\alpha}\partial_\rho\tilde{g}_{\nu\beta}. 
\label{com3}\end{align}

By substituting (\ref{com1,2}) and using the partial integration, the gravitational Lagrangian is
\begin{align}
\mathcal{L}_\text{gravity}&=\frac{1}{\kappa^2}\sqrt{-g}\big[R-2\Lambda\big] \label{gravity}\\
&=\frac{1}{\kappa^2}\big[\Omega^2\tilde{R}
+6\tilde{g}^{\mu\nu}\partial_\mu\Omega\partial_\nu\Omega-6H^2\Omega^4\big], \notag
\end{align}
where $\Lambda=3H^2$. Note that the Lagrangian density is defined including $\sqrt{-g}$ in this paper. 
In order to fix the gauge with respect to general coordinate invariance, 
we introduce the gauge fixing term \cite{Tsamis1992} 
\begin{align}
\mathcal{L}_\text{GF}&=-\frac{1}{2}a^{2}F_\mu F^\mu, \label{GF}\\
F_\mu&=\partial_\rho h_\mu^{\ \rho}-2\partial_\mu w+2h_\mu^{\ \rho}\partial_\rho\log a+4w\partial_\mu\log a. \notag
\end{align}
Henceforth the Lorentz indices are raised and lowered by $\eta^{\mu\nu}$ and $\eta_{\mu\nu}$ respectively.
The corresponding ghost term at the quadratic level is
\begin{align}
\mathcal{L}_\text{ghost}=&-a^2\partial^\nu\bar{b}^\mu 
\big\{{\eta}_{\mu\rho}\partial_\nu+\eta_{\nu\rho}\partial_\mu+2\eta_{\mu\nu}\partial_\rho(\log a)\big\}b^\rho\label{ghost}\\
&+\partial_\mu(a^2\bar{b}^\mu)
\big\{\partial_\nu+4\partial_\nu(\log a)\big\}b^\nu, \notag
\end{align}
where $b^\mu$ is the ghost field and $\bar{b}^\mu$ is the anti-ghost field. 

From (\ref{com3})-(\ref{ghost}), the quadratic part of the total gravitational Lagrangian density is
\begin{align}
\mathcal{L}_\text{quadratic}=a^4&\big[\ \frac{1}{2}a^{-2}\partial_\mu X\partial^\mu X-\frac{1}{4}a^{-2}\partial_\mu\tilde{h}^i_{\ j}\partial^\mu\tilde{h}^j_{\ i}
-a^{-2}\partial_\mu\bar{b}^i\partial^\mu b^i\label{quadratic}\\
&+\frac{1}{2}a^{-2}\partial_\mu h^{0i}\partial^\mu h^{0i}+H^2h^{0i}h^{0i}\notag\\
&-\frac{1}{2}a^{-2}\partial_\mu Y\partial^\mu Y-H^2Y^2\notag\\
&+a^{-2}\partial_\mu\bar{b}^0\partial^\mu b^0+2H^2\bar{b}^0b^0\big]. \notag
\end{align}
Here we have decomposed $h^i_{\ j},\ i,j=1, \cdots, 3$ into a trace part and a traceless part
\begin{align}
h^i_{\ j}=\tilde{h}^i_{\ j}+\frac{1}{3}h^k_{\ k}\delta^i_{\ j}=\tilde{h}^i_{\ j}+\frac{1}{3}h^{00}\delta^i_{\ j}.
\end{align} 
The action has been diagonalized by the following linear combination 
\begin{align}
X=2\sqrt{3}w-\frac{1}{\sqrt{3}}h^{00},\hspace{1em}Y=h^{00}-2w. 
\label{diagonalize}\end{align}
The quadratic action (\ref{quadratic}) contains two types of fields, 
massless and minimally coupled fields: $X,\tilde{h}^i_{\ j},b^i,\bar{b}^i$ 
and massless conformally coupled fields: $h^{0i},b^0,\bar{b}^0$, $Y$. 
We list the corresponding propagators as follows 
\begin{align}
\langle X(x)X(x')\rangle&=-\langle\varphi(x)\varphi(x')\rangle, \label{minimally}\\
\langle\tilde{h}^i_{\ j}(x)\tilde{h}^k_{\ l}(x')\rangle&=(\delta^{ik}\delta_{jl}+\delta^i_{\ l}\delta_j^{\ k}-\frac{2}{3}\delta^i_{\ j}\delta^k_{\ l})\langle\varphi(x)\varphi(x')\rangle, \notag\\
\langle b^i(x)\bar{b}^j(x')\rangle&=\delta^{ij}\langle\varphi(x)\varphi(x')\rangle, \notag
\end{align}
\begin{align}
\langle h^{0i}(x)h^{0j}(x')\rangle&=-\delta^{ij}\langle\phi(x)\phi(x')\rangle, \label{conformally}\\
\langle Y(x)Y(x')\rangle&=\langle\phi(x)\phi(x')\rangle, \notag\\
\langle b^0(x)\bar{b}^0(x')\rangle&=-\langle\phi(x)\phi(x')\rangle. \notag
\end{align}

Here $\varphi$ denotes a massless and minimally coupled scalar field and $\phi$ denotes a massless conformally coupled scalar field
\begin{align}
\langle\varphi(x)\varphi(x')\rangle&=\frac{H^2}{4\pi^2}\big\{\frac{1}{y}-\frac{1}{2}\log y+\frac{1}{2}\log a(\tau)a(\tau')+1-\gamma\big\}, 
\label{minimally0}\end{align} 
\begin{align}
\langle\phi(x)\phi(x')\rangle&=\frac{H^2}{4\pi^2}\frac{1}{y}, 
\label{conformally0}\end{align}
where $\gamma$ is Euler's constant and $y$ is the dS invariant distance
\begin{align}
y=\frac{-(\tau-\tau')^2+({\bf x}-{\bf x}')^2}{\tau\tau'}. 
\end{align}
It should be noted that the propagator for a massless and minimally coupled scalar field has 
the dS symmetry breaking logarithmic term: $\log a(\tau)a(\tau')$. 
To explain what causes the dS symmetry breaking, 
we recall the wave function for a massless and minimally coupled field
\begin{align}
\phi_{\bf p}(x)=\frac{H\tau}{\sqrt{2p}}(1-i\frac{1}{p\tau})e^{-ip\tau+i{\bf p}\cdot{\bf x}}. 
\end{align}
At the sub-horizon scale $P\equiv p/a(\tau)\gg H \Leftrightarrow p|\tau|\gg 1$, 
this wave function approaches to that in Minkowski space up to a cosmic scale factor
\begin{align}
\phi_{\bf p}(x)\sim\frac{H\tau}{\sqrt{2p}}e^{-ip\tau+i{\bf p}\cdot{\bf x}}. 
\end{align}
On the other hand, the behavior at the super-horizon scale $P\ll H$ is
\begin{align}
\phi_{\bf p}(x)\sim\frac{H}{\sqrt{2p^3}}e^{i{\bf p}\cdot{\bf x}}. 
\end{align}
The IR behavior indicates that the corresponding propagator has a scale invariant spectrum. 
As a direct consequence of it, the propagator has a logarithmic divergence from the IR contributions
in the infinite volume limit. 

To regularize the IR divergence, we introduce an IR cut-off $\varepsilon_0$ which fixes the minimum value of the comoving momentum. 
The minimum value of the physical momentum is  $\varepsilon_0/a(\tau )$ as their wavelength is stretched by cosmic expansion.
With this prescription, more degrees of freedom accumulate outside the cosmological horizon with cosmic evolution. 
Due to this increase, the propagator acquires the growing time dependence which spoils the dS symmetry. 
In tribute to its origin, we call this type of dS symmetry breaking term the IR logarithm. 
Physically speaking, $1/\varepsilon_0$ is recognized as an initial size of the Universe 
when the exponential expansion starts. 
For simplicity, we set $\varepsilon_0=H$ in (\ref{minimally0}). 

As there is explicit  time dependence in the propagator, 
physical quantities can acquire time dependence through the quantum loop corrections. 
We call them the quantum IR effects in dS space. 
We investigate such effects with respect to gauge couplings in this paper.
Before concluding this section, we introduce an approximation. 
Focusing on the leading IR effects, 
we can neglect conformally coupled modes of gravity since they do not induce the IR logarithm. 
By applying this approximation, we can identify the following two modes as
\begin{align}
h^{00}\simeq 2w\simeq\frac{\sqrt{3}}{2}X. 
\label{diagonalize1}\end{align}
From (\ref{minimally}) and (\ref{diagonalize1}), we have only to focus 
on the following propagators after retaining massless and minimally coupled modes from gravity
\begin{align}
\langle h^{00}(x)h^{00}(x')\rangle&\simeq-\frac{3}{4}\langle\varphi(x)\varphi(x')\rangle, \label{gravityp}\\
\langle h^{00}(x)h^i_{\ j}(x')\rangle&\simeq-\frac{1}{4}\delta^i_{\ j}\langle\varphi(x)\varphi(x')\rangle, \notag\\
\langle h^i_{\ j}(x)h^k_{\ l}(x')\rangle&\simeq(\delta^{ik}\delta_{jl}+\delta^i_{\ l}\delta_j^{\ k}-\frac{3}{4}\delta^i_{\ j}\delta^k_{\ l})\langle\varphi(x)\varphi(x')\rangle. \notag
\end{align}

%------------------------------------------------------------------------------------
\section{Soft graviton effects on gauge couplings}
\setcounter{equation}{0}

In the preceding section, we have reviewed that 
the propagators for some components of gravitation are time dependent in violation of the dS symmetry. 
By deriving the quantum equations of motion, 
we evaluate how the soft gravitons affect the gauge field dynamics. 
In \cite{Kahya2007,Miao2005(1),Miao2005(2),Miao2005(3),Giddings2010}, the effects of soft gravitons on 
the free matter fields at the super-horizon scale have been investigated. 
In contract, we focus on the soft graviton effects on the local field dynamics at the sub-horizon scale which are directly observable. 
In this section we investigate the soft gravitational effects on the local dynamics of gauge theories. 

The action of the gauge theory is given by 
\begin{align}
S_\text{gauge}&=\int\sqrt{-g}d^4x\big[-\frac{1}{4g^2}g^{\mu\rho}g^{\nu\sigma}F_{\mu\nu}^a F_{\rho\sigma}^a
+i\bar{\psi}e^\mu_{\ \alpha}\gamma^\alpha D_\mu\psi\big], \label{gauge}\\
&=\int d^4x\big[-\frac{1}{4g^2}\tilde{g}^{\mu\rho}\tilde{g}^{\nu\sigma}F_{\mu\nu}^a F_{\rho\sigma}^a
+i\bar{\psi}\tilde{e}^\mu_{\ \alpha}\gamma^\alpha \tilde{D}_\mu\psi\big], \notag
\end{align}
where $e^\mu_{\ \alpha}$ is the vierbein 
\begin{align}
e^\mu_{\ \alpha}=\Omega^{-1}\tilde{e}^\mu_{\ \alpha},\hspace{1em}
\tilde{e}^\mu_{\ \alpha}=(e^{-\frac{\kappa}{2}h})^\mu_{\ \alpha}. 
\end{align}
We have redefined the Dirac field: $\Omega^\frac{3}{2}\psi\to\psi$ in the second line of (\ref{gauge}).   
We consider that the generic gauge group and the Dirac field could be in any representation, 
\begin{align}
F_{\mu\nu}^a&=\partial_\mu A^a_\nu-\partial_\nu A^a_\mu+f^{abc}A^b_\mu A^c_\nu, 
\hspace{1em}[t^a,t^b]=if^{abc}t^c, \\
\tilde{D}_\mu&\simeq\partial_\mu-iA^a_\mu t^a. \notag
\end{align}
Strictly speaking, the covariant derivative $\tilde{D}_\mu$ contains the spin connection made of $\tilde{e}^\mu_{\ \alpha}$ but we have neglected it.  
It is because the spin connection consists of differentiated gravitational fields. 
When we consider the gauge field at the sub-horizon scale $P\gg H$, 
the kinetic term is dominant in comparison with the terms where the IR logarithms from gravitational fields are differentiated: 
\begin{align}
P\gg H\ \Rightarrow\ \log a(\tau)\partial\gg \partial\log a(\tau). 
\label{sub}\end{align}
In other words, $\log a(\tau )$ factor can be regarded as a constant since it varies slowly.
This is the reason why we can absorb the IR logarithm in front of the kinetic term by the wave function renormalization
of the fields at the one-loop level \cite{Kitamoto2012}.

To obtain the quantum equation of the gauge field, we separate the field components into the classical field and the quantum fluctuation 
\begin{align}
A_\mu\to \hat{A}_\mu+A_\mu,\hspace{1em}\psi\to\hat{\psi}+\psi. 
\end{align}
In addition, we introduce the following gauge fixing term
\begin{align}
\mathcal{L}_\text{gf}=-\frac{\xi}{2}\sqrt{-g}\big\{\frac{1}{\sqrt{-g}}\hat{D}_\mu({\sqrt{-g}g^{\mu\nu}A_\nu})\big\}^2, 
\end{align}
where $\xi$ is the gauge parameter and $\hat{D}_{\mu}$ is the covariant derivative with respect to the classical field. 
We adopt the background gauge which preserves manifest gauge invariance with respect to the classical field.
Up to the quadratic action, the bosonic action in the Feynman gauge $\xi=1$ is
\begin{align}
\frac{1}{g^2}\int d^4x\ \tilde{g}^{\mu\rho}\tilde{g}^{\nu\sigma}\big[-\frac{1}{4}\hat{F}_{\mu\nu}^a \hat{F}_{\rho\sigma}^a
-\frac{1}{2}(\hat{D}_{\mu}A_{\nu})^a(\hat{D}_{\rho}A_{\sigma})^a
-\hat{F}_{\mu\nu}^a A_{\rho}^bA_{\sigma}^cf_{abc}
+(\hat{D}_{\rho}\hat{F}_{\mu\nu})^aA_{\sigma}^a
\big]. 
\label{EEoMp}
\end{align}
We show that our results do not depend on the parameter $\xi$ in the subsequent investigation. 

In investigating interacting field theories on a time dependent background like dS space, 
we need to adopt the Schwinger-Keldysh path \cite{Schwinger1961,Keldysh1964} 
\begin{align}
&\parbox{\SKpathl}{\usebox{\SKpath}}, \label{SKpath}\\
&\hspace{3em}\int_C dt = \int^\infty_{-\infty} dt_+ - \int^\infty_{-\infty} dt_-. \notag
\end{align}
Since there are two time indices $+,-$ in this path, the propagator has four components 
\begin{align}
\begin{pmatrix} \langle\varphi_+(x)\varphi_+(x')\rangle & \langle\varphi_+(x)\varphi_-(x')\rangle \\
\langle\varphi_-(x)\varphi_+(x')\rangle & \langle\varphi_-(x)\varphi_-(x')\rangle \end{pmatrix}
=\begin{pmatrix} \langle T\varphi(x)\varphi(x')\rangle & \langle\varphi(x')\varphi(x)\rangle \\
\langle\varphi(x)\varphi(x')\rangle & \langle\tilde{T}\varphi(x)\varphi(x')\rangle \end{pmatrix}. 
\label{4propagators}\end{align}
Here $\varphi$ denotes the quantum fluctuation of an arbitrary field component and $\tilde{T}$ denotes the anti-time ordering. 
By dealing with the quantum fluctuations along the path (\ref{SKpath}), 
we derive the effective gauge field equation \cite{Hu1997}
\begin{align}
\frac{\delta \Gamma[(\hat{A}_\mu)_+,(\hat{A}_\mu)_-,\hat{\psi}_+,\hat{\psi}_-]}{\delta (\hat{A}_\mu)_+(x)}
\Big|_{(\hat{A}_\mu)_+=(\hat{A}_\mu)_-=\hat{A}_\mu,\ \hat{\psi}_+=\hat{\psi}_-=\hat{\psi}}=0, 
\label{EEoM}\end{align}
where $\Gamma$ denotes the effective action. 

Up to the one-loop level and $\mathcal{O}(\log a(\tau))$, the bosonic quantum equation of motion is evaluated as 
\begin{align}
&\frac{1}{g^2}(\hat{D}_\mu \hat{F}^{\mu\nu})^a(x)
+\kappa^2\big\{\langle (h^{\mu\rho})_+(x)(h^{\nu\sigma})_+(x)\rangle
+\frac{1}{2}\langle (h^{\mu\alpha})_+(x)(h_\alpha^{\ \rho})_+(x)\rangle\eta^{\nu\sigma} \label{F1}\\
&\hspace{9em}+\frac{1}{2}\eta^{\mu\rho}\langle (h^{\nu\alpha})_+(x)(h_\alpha^{\ \sigma})_+(x)\rangle\big\}\frac{1}{g^2}(\hat{D}_\mu \hat{F}_{\rho\sigma})^a(x). \notag
\end{align} 
Here we have neglected the contribution from the linear term in the gauge field fluctuation $A^a_\mu$ in the action (\ref{EEoMp})
as it can be eliminated by a judicious redefinition of the quantum field $A^a_{\mu}$ by the metric degrees of freedom. 
In this way, the gauge parameter $\xi$ does not influence (\ref{F1}) 
since there is no contribution from the propagator of the gauge field. 

As explained in the previous section, 
we focus on the massless and minimally coupled modes of gravitational field. 
By adopting an ultra-violet (UV) regularization, 
the propagator at the coincident point is estimated as follows
\begin{align}
\langle\varphi(x)\varphi(x)\rangle=\text{(UV divergent const)}+\frac{H^2}{4\pi^2}\log a(\tau). 
\end{align}
In the subsequent investigation, we focus on the time dependent dS symmetry breaking part 
\begin{align}
\langle\varphi(x)\varphi(x)\rangle\simeq\frac{H^2}{4\pi^2}\log a(\tau). 
\label{minimallyc}\end{align}
By substituting (\ref{gravityp}) and (\ref{minimallyc}), (\ref{F1}) is evaluated as
\begin{align}
\frac{1}{g^2}\big\{1+\frac{3\kappa^2H^2}{8\pi^2}\log a(\tau)\big\}(\hat{D}_\mu \hat{F}^{\mu\nu})^a(x). 
\end{align}
The result preserves the gauge symmetry manifestly 
and indicates that the coupling of the gauge interaction decreases with cosmic expansion 
\begin{align}
g_\text{eff}\simeq g\big\{1-\frac{3\kappa^2H^2}{16\pi^2}\log a(\tau)\big\}. 
\label{EGc}\end{align}
It is because there is no wave function renormalization for the classical gauge field in the background gauge.
The Lorentz invariance is also preserved and the velocity of light is not renormalized just like in the massless 
scalar and Dirac field cases \cite{Kitamoto2012}. 

We should emphasize that the screening of the coupling (\ref{EGc}) is not a renormalization group flow. 
The renormalization group flow indicates logarithmic dependence of momentum scale which comes from quantum contribution from physical momentum scale $P=p/a(\tau)$ to UV cut-off $\Lambda_\text{UV}$. 
Then it is constant for a fixed physical momentum scale and respects the dS symmetry. 
In contrast, the screening (\ref{EGc}) manifestly breaks the dS symmetry. 
The time dependence originates in the gravitational fluctuation at the super-horizon scale. 

We further investigate the fermionic current $\hat{\bar{\psi}}\gamma^\nu t^a\hat{\psi}$ in the quantum equation of motion.
Up to the one-loop level, this term is evaluated as
\begin{align}
\hat{\bar{\psi}}(x)\gamma^\nu t^a\hat{\psi}(x)
&+\frac{\kappa^2}{8}\langle (h^\nu_{\ \rho})_+(x)(h^\rho_{\ \alpha})_+(x)\rangle\hat{\bar{\psi}}(x)\gamma^\alpha t^a\hat{\psi}(x) \label{psi2}\\
&+\frac{\kappa^2}{4}\hat{\bar{\psi}}(x)\int d^4x'\ c_{AB}\langle(h^\nu_{\ \alpha})_+(x)(h^\rho_{\ \beta})_A(x')\rangle\gamma^\alpha t^a
\langle\psi_+(x)\partial_\rho'\bar{\psi}_B(x')\rangle\gamma^\beta\hat{\psi}(x') \notag\\
&-\frac{\kappa^2}{4}\int d^4x'\ c_{AB}\langle(h^\nu_{\ \alpha})_+(x)(h^\rho_{\ \beta})_A(x')\rangle\hat{\bar{\psi}}(x')\gamma^\beta
\langle\partial_\rho'\psi_B(x')\bar{\psi}_+(x)\rangle\gamma^\alpha t^a\hat{\psi}(x), \notag\\
&-\frac{\kappa^2}{4}\int d^4x'd^4x''\ c_{AB}c_{CD}\langle(h^\rho_{\ \alpha})_A(x')(h^\sigma_{\ \beta})_C(x'')\rangle \notag\\
&\hspace{3em}\times\hat{\bar{\psi}}(x')\gamma^\alpha\langle\partial'_\rho\psi_B(x')\bar{\psi}_+(x)\rangle\gamma^\nu t^a
\langle\psi_+(x)\partial_\sigma''\bar{\psi}_D(x'')\rangle\gamma^\beta\hat{\psi}(x''), \notag
\end{align}
where $A,B,C,D$ are assigned by the Schwinger-Keldysh indices $+,-$ and the matrix $c_{AB}$ is 
\begin{align}
c_{AB}=\begin{pmatrix} 1 & 0 \\ 0 & -1 \end{pmatrix}. 
\end{align}

By substituting the identity
\begin{align}
\langle\psi(x)\bar{\psi}(x')&=i\gamma^\mu\partial_\mu\langle\phi(x)\phi(x')\rangle, \label{Dp}\\
\langle\phi(x)\phi(x')&=\frac{1}{4\pi^2\Delta x^2},\hspace{1em}\Delta x^\mu\equiv x^\mu-x'^\mu, \notag
\end{align}
(\ref{psi2}) are represented by the integrals involving massless and minimally coupled fields and a massless conformally coupled field. 
Recall that we have redefined the Dirac field $\Omega^\frac{3}{2}\psi\to\psi$. 
To evaluate these integrals up to $\mathcal{O}(\log a(\tau))$, we may take the coincident limit of the propagator for massless minimally coupled fields: 
\begin{align}
\langle (h^\nu_{\ \alpha})_+(x)(h^\rho_{\ \beta})_+(x)\rangle\int d^4x',\hspace{1em}
\langle(h^\rho_{\ \alpha})_+(x)(h^\sigma_{\ \beta})_+(x)\rangle\int d^4x'd^4x''. 
\label{out}\end{align}
This approximate prescription has been introduced 
in the investigations of Yukawa theory and scalar QED \cite{Duffy2005,Woodard2006,Prokopec2007}. 
In this approximation (\ref{out}), only the following local term contributes to the remaining integral 
\begin{align}
\langle\partial_\mu\partial_\nu\phi_+(x)\phi_+(x')\rangle
=-\langle\partial_\mu\phi_+(x)\partial_\nu'\phi_+(x')\rangle
\to -i\delta_\mu^{\ 0}\delta_\nu^{\ 0}\delta^{(4)}(x-x'). 
\label{local}\end{align}
From (\ref{Dp})-(\ref{local}), (\ref{psi2}) is evaluated as
\begin{align}
\big\{1+\frac{3\kappa^2H^2}{128\pi^2}\log a(\tau)\big\}\hat{\bar{\psi}}(x)\gamma^\nu t^a\hat{\psi}(x). 
\label{nr}\end{align}
We recall here that the IR logarithm due to soft gravitons modifies the kinetic term of the Dirac field. 
We have shown that this change of the kinetic term can be absorbed by the following wave function renormalization \cite{Kitamoto2012} 
\begin{align}
\psi\to Z_\psi\psi,\hspace{1em}Z_\psi\simeq\big\{1-\frac{3\kappa^2H^2}{256\pi^2}\log a(\tau)\big\}. 
\label{Z}\end{align}
It should be noted that the quantum correction in (\ref{nr}) can also be absorbed by the identical wave function renormalization. 
We therefore conclude that the fermionic current is not renormalized by soft gravitons after the wave function renormalization.
The non-renormalization of the fermionic current is consistent with (\ref{EGc}) since the fermionic current  
$\hat{\bar{\psi}}\gamma^\nu t^a\hat{\psi}$ in the quantum equation of motion does not involve the coupling of the gauge interaction. 

As for soft gravitational effects in a free Dirac field, we should mention the previous studies \cite{Miao2005(1),Miao2005(2),Miao2005(3)}. 
The investigations are performed in the same gauge, but in a different parametrization of the metric: 
\begin{align}
g_{\mu\nu}=a^2(\tau)(\eta_{\mu\nu}+2\kappa\Phi(x)\eta_{\mu\nu}+\kappa\Psi_{\mu\nu}(x)),\hspace{1em}\eta^{\mu\nu}\Psi_{\mu\nu}=0, 
\end{align}
and with a different matter field redefinition: $a^\frac{3}{2}\psi\to\psi$. 
The authors of these papers derive the solution of the effective field equation at the super-horizon scale. 
In contrast, we investigate the off-shell effective field equation at the sub-horizon scale. 
The existence of nonzero virtuality indicates that only the local terms in quantum equation contribute to the dS symmetry breaking. 
While they consider the full modes of gravity and renormalize UV divergences, 
we neglect differentiated gravitational fields and non-minimally coupled modes of gravity. 
Furthermore we extract the logarithmic term of the propagator for minimally coupled modes. 

As far as we focus on the sub-horizon dynamics with the dS symmetry breaking logarithms, the approximation approach is consistent with the exact one although their results are different. 
It is due to different metric parametrization and the choice of fundamental field in both calculations \cite{Kitamoto2012}. 
We will report the prescription to compensate these quantization scheme dependences \cite{Parameter}. 

As seen in (\ref{EGc}), the gravitational fluctuation at the super-horizon scale influences the local dynamics of the gauge field at the sub-horizon scale. 
The identity (\ref{EGc}) holds for a generic gauge group. 
Here we have evaluated the effective coupling up to the one-loop level. 
By the power counting of the IR logarithms, 
the leading IR effect at the $n$-loop level is estimated as of order $(\kappa^2H^2\log a(\tau))^n$. 
In order to investigate the physics at late times $\kappa^2H^2\log a(\tau)\sim 1$, 
we need a non-perturbative method. 
Aside from this open problem, 
we examine the gauge dependences of the perturbative results in the next section. 

%------------------------------------------------------------------------------------
\section{Gauge dependence}
\setcounter{equation}{0}

In the preceding sections, we have investigated the IR effects with the gauge fixing term (\ref{GF}).  
To check whether the obtained results are physical, 
we need to investigate the gauge dependence of them. 
Here we adopt the following gauge fixing term with a parameter $\beta$: 
\begin{align}
\mathcal{L}_\text{GF}&=-\frac{1}{2}a^2F_\mu F^\mu, \label{beta}\\
F_\mu&=\beta\partial_\rho h_\mu^{\ \rho}-2\beta\partial_\mu w+\frac{2}{\beta}h_\mu^{\ \rho}\partial_\rho\log a+\frac{4}{\beta}w\partial_\mu\log a. \notag
\end{align}
This gauge fixing term coincides with (\ref{GF}) at $\beta=1$. 
For simplicity, we consider the case $|\beta^2-1|\ll 1$ 
where the deviation from (\ref{GF}) can be investigated perturbatively. 
The deformation of the action at $\mathcal{O}(\beta^2-1)$ is 
\begin{align}
\delta \mathcal{L}_{\beta^2-1}\simeq-\frac{1}{2}(\beta^2-1)a^2&\big[\ 
\partial_\mu h^{00}\partial^\mu h^{00}-3\partial_0 h^{00}\partial_0 h^{00}-\frac{5}{9}\partial_i h^{00}\partial_i h^{00}\label{beta1}\\
&-\frac{4}{3}\partial_i h^{00}\partial_k \tilde{h}^{ki}+\partial_k\tilde{h}^k_{\ i}\partial_l\tilde{h}^{li}\big]. 
\notag
\end{align}
Here we have neglected massless conformally coupled modes. 
In addition, we have ignored ghost fields since they do not couple to matter fields. 

We have found that the additional term (\ref{beta1}) modifies the gravitational propagator as follows \cite{Kitamoto2012}
\begin{align}
\langle (h^{\mu\nu})_+(x)(h^{\rho\sigma})_+(x)\rangle
\to\big\{1-(\beta^2-1)\big\}\langle (h^{\mu\nu})_+(x)(h^{\rho\sigma})_+(x)\rangle. 
\label{bgravityp'}\end{align}
From this fact, we can conclude that the IR effect to the fermionic current $\hat{\bar{\psi}}\gamma^\nu t^a\hat{\psi}$ can be absorbed 
also by the wave function renormalization in the gauge (\ref{beta}): 
\begin{align}
\big\{1+(2-\beta^2)\frac{3\kappa^2H^2}{128\pi^2}\log a(\tau)\big\}\bar{\hat{\psi}}\gamma^\nu t^a\hat{\psi}, 
\label{betanr}\end{align}
\begin{align}
\psi\to Z_\psi\psi,\hspace{1em}Z_\psi&\simeq 1-(2-\beta^2)\frac{3\kappa^2H^2}{256\pi^2}\log a(\tau). 
\label{betaZ}\end{align}
So the soft gravitons are also found to screen the gauge coupling in this gauge as
\begin{align}
g_\text{eff}\simeq g\big\{1-(2-\beta^2)\frac{3\kappa^2H^2}{16\pi^2}\log a(\tau)\big\}. 
\label{betaEGc}\end{align}
We find that the effective gauge coupling depends on the gauge parameter. 

To give a physical interpretation to this result, 
we recall our previous results concerning $\lambda_4\phi^4$ and $\lambda_Y\phi\bar{\psi}\psi$ interactions \cite{Kitamoto2012}. 
Soft gravitons also screen these interactions with cosmic expansion. 
In the gauge (\ref{beta}), these couplings evolve as follows 
\begin{align}
(\lambda_4)_\text{eff}&\simeq \lambda_4\big\{1-(2-\beta^2)\frac{21\kappa^2H^2}{16\pi^2}\log a(\tau)\big\}, \label{betaEcs}\\
(\lambda_Y)_\text{eff}&\simeq \lambda_Y\big\{1-(2-\beta^2)\frac{39\kappa^2H^2}{128\pi^2}\log a(\tau)\big\}. \notag
\end{align}
The effective couplings (\ref{betaEGc}), (\ref{betaEcs}) can be represented 
by the following common function with definite scaling exponents: 
\begin{align}
(\lambda_4)_\text{eff}=\lambda_4 f(\tau)^\frac{21}{4},\hspace{1em}
(\lambda_Y)_\text{eff}=\lambda_Y f(\tau)^\frac{39}{32},\hspace{1em}
g_\text{eff}=g f(\tau)^\frac{3}{4}, 
\end{align}
where $f(\tau)=1-(2-\beta^2)\frac{\kappa^2H^2}{4\pi^2}\log a(\tau)$. 
Out of these couplings we can form gauge independent ratios as follows
\begin{align}
(\lambda_Y)_\text{eff}/\lambda_Y=\big\{(\lambda_4)_\text{eff}/\lambda_4\big\}^\frac{13}{56},\hspace{1em}
g_\text{eff}/g=\big\{(\lambda_4)_\text{eff}/\lambda_4\big\}^\frac{1}{7}. 
\label{rs}\end{align}

We interpret our findings as follows. 
The time dependence of each effective coupling is gauge dependent 
since there is no unique way to specify the time as it depends on an observer. 
A sensible strategy may be to pick a particular coupling and use its time evolution as a physical time. 
In (\ref{rs}), the coupling of the quartic interaction has been assigned to this role. 

In this setting the relative scaling exponents measure the time evolution of the couplings in terms of a physical time. 
We should emphasize that up to the one-loop level, they are just numbers which do not depend on initial time. 
Although the choice of time is not unique, the relative scaling exponents are gauge independent and well defined. 
The situation is analogous to the scaling exponents of the local operators in two dimensional quantum gravity. 
The scaling exponents of the individual operators are gauge dependent 
since there is no unique way to specify the scale 
but the relative scaling exponents are gauge invariant \cite{KN,KKN}. 
We need to mention here that our investigation is performed assuming $|\beta^2-1|\ll 1$. 
We still need to check the validity of this picture against large deformations of the gauge parameter. 

%------------------------------------------------------------------------------------
\section{Conclusion}
\setcounter{equation}{0}

The gravitational fields contain almost scale invariant modes
in the exponentially expanding space-time such as the inflation era or present Universe. 
The loop contribution is sensitive to the degrees of freedom at the super-horizon scale
and could give rise to dS symmetry breaking effects. 

In this paper, we have investigated the effects of soft gravitons 
on the local dynamics of gauge field theories up to the one-loop level. 
Along with our previous results in $\phi^4$ and Yukawa theories \cite{Kitamoto2012}, 
the effective couplings of the gauge theories decrease with cosmic evolution. 
The screening effect is universal independent of the gauge group. 
We may chose any coupling as a physical time. 
Thus the relative scaling exponents of the couplings are expected to be gauge independent. 
In fact we have checked their gauge independence under an infinitesimal change of gauge. 
The check of the independence against large deformations is left as an open problem. 

Since $\kappa^2H^2\ll 1$ in most cases of physical interest, 
the quantum effects from gravity seem to be suppressed by it. 
However we have found that the effective couplings are associated with the growing time dependence: $\log a(\tau)$. 
This factor enhances the quantum effects with cosmic evolution. 
Due to excessive smallness of the Hubble parameter in the present Universe: $\kappa^2H^2\sim 10^{-120}$, 
we cannot observe such enhancements now but it is relevant to the ultimate fate of the Universe.  
In the inflation era, the IR effects are much larger as $\kappa^2H^2$ could be as large as $10^{-10}$. 
The enhancement depends on how long the inflation era continues.  

By the power counting of the IR logarithms, 
the leading IR effects at the $n$-loop level are estimated as $(\kappa^2H^2\log a(\tau))^n$. 
So in order to investigate the eventual contributions to physical quantities, 
we need to evaluate the IR effects nonperturbatively. 
Although the IR effects from specific matter fields have been investigated nonperturbatively 
\cite{Starobinsky1994,Woodard2005,Woodard2006,Prokopec2007,Kitamoto2011}, 
the nonperturbative approach for the IR effects from gravitons is an open issue.

Our investigations in this paper and \cite{Kitamoto2012} have clarified the IR effects 
on all dimensionless couplings from soft gravitons up to the one-loop level. 
These investigations are valid to describe the local dynamics at the initial stage $\kappa^2H^2\log a(\tau)\ll 1$. 
Since the cosmological constant is a function of the couplings of the microscopic theory, 
it may acquire time dependence if the couplings evolve with time. 
So in addition to pure matter and gravity contributions separately, 
we need to investigate such effects to understand possible time dependence of the cosmological constant. 
The mixed matter and gravity contribution starts at the two-loop level. 

We have shown that the leading IR effects due to the soft gravitons of the super-horizon scale
can be absorbed into the renormalization of the fields and couplings inside
the cosmological horizon. We believe that this result has an implication for the
super-horizon soft graviton loop effect on the cosmic microwave background.
Let us consider the density fluctuation due to scalar modes.
The standard argument relates it to the two point function of the minimally coupled scalar field (inflaton) at the
horizon crossing as it is identified with the curvature perturbation which remains constant
after the horizon crossing in the comoving gauge \cite{Mukhanov1990,Maldacena2002}.
Our investigation shows that the leading IR effects on the scalar two point functions due to soft gravitons
are absent as they can be renormalized away. The renormalization of the couplings modifies slow-roll
parameters $\epsilon$ and $\eta$. However such modification cancels for the both quadratic and quartic
potential. The remaining issue is to clarify whether the curvature perturbation receives leading IR effects
after the horizon crossing.

\section*{Acknowledgment}
This work is supported in part by the Grant-in-Aid for Scientific Research
from the Ministry of Education, Science and Culture of Japan.

%-------------------------------------------------------------------------------------

\end{document}